\documentclass[aps,prd,onecolumn,eqsecnum,amsmath,nofootinbib,preprintnumbers]{revtex4}%

\usepackage{color,graphicx,float,subfigure,xcolor}
\usepackage{amsfonts,amssymb,theorem,mathrsfs,times}
\usepackage{bm}
\usepackage{multirow}
\usepackage{mathtools}
\usepackage{amsfonts,amssymb,theorem,mathrsfs}
\usepackage{dsfont}
\usepackage{setspace}
\usepackage{amsmath} 
\usepackage[colorlinks,linkcolor=blue,anchorcolor=blue,citecolor=blue]{hyperref}   
\usepackage{ulem}   
\usepackage[english]{babel}
\usepackage{physics}

{\theorembodyfont{\upshape}
	}
{\theorembodyfont{\upshape}
	}
{\theorembodyfont{\upshape}
	}
{\theorembodyfont{\upshape}
	}
{\theorembodyfont{\upshape}
	}
{\theorembodyfont{\upshape}
	}
\addtocounter{MaxMatrixCols}{10}
\newcommand{\dalm}{\kern1pt\vbox{\hrule height 0.9pt\hbox{\vrule width 0.9pt\hskip 2.5pt\vbox{\vskip 5.5pt}\hskip 3pt\vrule width 0.3pt}\hrule height 0.3pt}\kern1pt}

\begin{document}
\thispagestyle{empty}
\title{Spectrum instability and greybody factor stability for parabolic approximation of Regge-Wheeler potential}
	
%



\author{Libo Xie$^{a\, ,b\, ,c}$\footnote{e-mail address: xielibo23@mails.ucas.ac.cn}}

\author{Liang-Bi Wu$^{a\, ,c}$\footnote{e-mail address: liangbi@mail.ustc.edu.cn (corresponding author)}}

\author{Zong-Kuan Guo$^{b\, ,a\, ,c}$\footnote{e-mail address: guozk@itp.ac.cn}}


	
\affiliation{${}^a$School of Fundamental Physics and Mathematical Sciences, Hangzhou Institute for Advanced Study, UCAS, Hangzhou 310024, China}

\affiliation{${}^b$CAS Key Laboratory of Theoretical Physics, Institute of Theoretical Physics, Chinese Academy of Sciences, Beijing 100190, China}

\affiliation{${}^c$University of Chinese Academy of Sciences, Beijing 100049, China}


\date{\today}
	
\begin{abstract}
We investigate the stability of QNM spectra and greybody factors in the Schwarzschild black hole by approximating the Regge-Wheeler potential with a piecewise parabolic form and treating the deviation as a perturbation. We find that QNM spectra are sensitive to small perturbations, while greybody factors remain stable. This piecewise parabolic approximated potential gives rise to the long-lived modes whose imaginary parts remain close to zero and decrease slowly with overtone number increasing. The reflection coefficient shows distinct resonance feature in the high-frequency regime that are absent in the original R-W case. For the calculation of greybody factors, we employ an analytic method based on transfer matrix technique, and this approach can also be effectively used in other effective potential cases.
\end{abstract}

\maketitle
	
\section{Introduction}
Quasinormal modes (QNMs) characterize the intrinsic oscillatory response of black holes to external  perturbations. These modes are described by a discrete set of complex frequencies, with the real part corresponding to the oscillation frequency and the imaginary part determining the decay rate. In the aftermath of compact object mergers, such as a binary black hole system, the newly formed remnant black hole undergoes a relaxation process known as the ringdown phase, during this period it emits gravitational radiation dominated by QNMs and these damped oscillations are only determined by the geometry of the final black hole spacetime. In this way, QNMs serve not only as spectral fingerprints of black holes, but also provide a powerful tool for testing the Kerr hypothesis and exploring the nature of gravity in the strong field regime~\cite{Berti:2009kk,Konoplya:2011qq,Bolokhov:2025uxz,Rao:2025rop}.

The QNM spectra of black holes exhibit an instability~\cite{Konoplya:2022pbc,Jaramillo:2020tuu}. In other words, these spectra will shift disproportionately far in the complex plane in response to seemingly minor external perturbations. Initial studies of QNM spectrum instability were provided by Nollert and Price~\cite{Nollert:1996rf,Nollert:1998ys}. There are two main categories of methods used to study spectrum instability in the aspect of QNMs. The first approach involves modifying the effective potential based on various reasons, which can be either physically motivated or artificially constructed~\cite{Qian:2020cnz,Daghigh:2020jyk,Liu:2021aqh,Li:2024npg,Berti:2022xfj,Cheung:2021bol,Yang:2024vor,Courty:2023rxk,Cardoso:2024mrw,Ianniccari:2024ysv,MalatoCorrea:2025iuc}. Furthermore, based on the smoothness of the modified effective potential, the above studies can be further classified. On the one hand, Refs. \cite{Qian:2020cnz,Daghigh:2020jyk,Liu:2021aqh,Li:2024npg} focus on non-smooth modifications of the effective potential to study the instability of QNM spectra. An alternative mechanism of the black hole echoes can be found for the discontinuous effective potential~\cite{Liu:2021aqh}. On the other hand, Refs. \cite{Berti:2022xfj,Cheung:2021bol,Yang:2024vor,Courty:2023rxk,Cardoso:2024mrw,Ianniccari:2024ysv,MalatoCorrea:2025iuc} investigate the instability of QNM spectra by adding small bump to the original effective potential. While finding a physically meaningful interpretation for such \textit{ad hoc} modifications in an astrophysical context can be challenging, such small bump in the effective potential of wave equation can be interpreted within the framework of the analogue black hole models~\cite{MalatoCorrea:2025iuc}.

Pseudospectrum analysis~\cite{trefethen2020spectra}, as a second approach, is utilized to investigate the instability of the QNM spectrum~\cite{Boyanov:2024fgc, Jaramillo:2020tuu,Destounis:2023ruj,Jaramillo:2021tmt}. This approach focuses on examining the characteristic properties of non-self-adjoint operators in dissipative systems, employing visual methods to elucidate the instabilities within their spectral configurations. In the context of gravity theory, pseudospectra have been employed to identify qualitative characteristics that serve as indicators of spectrum instability across a range of spacetimes, including asymptotically flat black holes~\cite{Jaramillo:2020tuu,Destounis:2021lum,Cao:2024oud,Cao:2024sot}, asymptotically AdS black holes or black branes~\cite{Arean:2024afl,Garcia-Farina:2024pdd,Arean:2023ejh,Boyanov:2023qqf,Cownden:2023dam,Carballo:2025ajx}, asymptotically dS black holes~\cite{Sarkar:2023rhp,Destounis:2023nmb,Luo:2024dxl,Warnick:2024usx}, horizonless compact objects~\cite{Boyanov:2022ark}, and Rezzolla-Zhidenko metric~\cite{Siqueira:2025lww}. Transient dynamics related to pseudospectra are studied in~\cite{Carballo:2024kbk,Jaramillo:2022kuv,Chen:2024mon,Carballo:2025ajx}. Furthermore, regarding the rotational situation, the pseudospectrum analysis of analogous black hole is presented in~\cite{dePaula:2025fqt}, while the pseudospectrum analysis of Kerr black holes is provided by~\cite{Cai:2025irl}.

However, it was proposed that the black hole greybody factors are more robust observables~\cite{Oshita:2023cjz,Rosato:2024arw,Oshita:2024fzf,Ianniccari:2024ysv,Wu:2024ldo} unlike the QNM spectra. The greybody factor is a fundamental concept in black hole physics and gravitational wave research, with significant applications to understanding gravitational wave~\cite{Oshita:2023cjz}. Both QNMs and greybody factors are characteristic quantities that reflect the geometric structure of spacetime, just with different boundary conditions. Their correspondence first established in spherically symmetric black hole by using WKB method~\cite{Konoplya:2024lir}. This finding has stimulated considerable interest in the study of the greybody factor and QNMs. Recently, there are many studies on the greybody factor and QNMs. The correspondence of them is later extended to the rotating case~\cite{Konoplya:2024vuj}, symmetric traversable wormholes~\cite{Bolokhov:2024otn}, quantum corrected black holes~\cite{Skvortsova:2024msa}, and massive ﬁelds in Schwarzschild-de Sitter spacetime~\cite{Malik:2024cgb}. Ref. \cite{Lutfuoglu:2025ohb} investigates the QNMs and greybody factors of axial gravitational perturbations by modeling the eﬀective quantum corrections through an anisotropic ﬂuid energy-momentum tensor. Subsequently, the applicability and limitations of the QNM-GBF correspondence have been discussed in~\cite{Pedrotti:2025upg}. Furthermore, other robust physical observable quantities, scattering cross section and the absorption cross section, are constructed from the Regge poles~\cite{Torres:2023nqg}. More recently, Regge poles and absorption cross sections with several modified Regge-Wheeler potentials ($C^0$ discontinuity) are studied~\cite{Li:2025ljb}.

The greybody factor will change both due to changes in boundary conditions~\cite{Rosato:2025byu} and due to changes in effective potential~\cite{Rosato:2024arw,Oshita:2024fzf,Ianniccari:2024ysv,Wu:2024ldo}. While previous investigations of greybody factors predominantly addressed scenarios involving smooth variations of the effective potential, the present study focuses on characterizing its behavior under non-smooth variations. This work employs a piecewise parabolic approximation as perturbation of the R-W potential ($C^1$ discontinuity) to systematically examine the stability of QNMs spectra and greybody factors (Note that the instability of QNM spectra for the piecewise linear approximation is studied in~\cite{Daghigh:2020jyk}).

The paper is organized as follows. In Sec. \ref{sec: parabolic potential}, we demonstrate how the R-W potential can be approximated by a piecewise parabolic potential. In Sec. \ref{stability_QNMs}, we use the multi-domain spectral collocation method to solve the QNM problem for the piecewise parabolic potential. In Sec. \ref{sec: GFs}, the greybody factor is derived and compared with the one under the R-W potential. Sec. \ref{sec: conclusions} is the conclusions and discussion. In addition, there are three appendices. Appendix \ref{solution_parabolic_potential} shows the analytical solution for a parabolic potential. Appendix \ref{construction_M} gives the construction of the matrix $\mathbf{M}(\omega)$, which is derived from the matching conditions at the interface of adjacent intervals. In Appendix \ref{app: greybody factor}, we obtain two analytic expressions of amplitude functions $A^{\text{in}}(\omega)$ and $A^{\text{out}}(\omega)$ for the piecewise parabolic potential by using the transfer matrix formalism. 

\section{The piecewise parabolic approximation of the Regge-Wheeler potential}\label{sec: parabolic potential}
In black hole perturbation theory, axial perturbations in the time-domain on a Schwarzschild black hole are governed by the Regge-Wheeler (R-W) equation:
\begin{eqnarray}\label{Regge_Wheeler_equation}
   \Big(\frac{\partial^2}{\partial t^2}- \frac{\partial^2}{\partial x^2}+ V_{\text{R-W}}(r(x))\Big)\Psi(t, x) = 0\, ,
\end{eqnarray}
where $V_{\text{R-W}}(r)$ represents the Regge-Wheeler potential, given by
\begin{eqnarray}\label{RW_potential}
   V_{\text{R-W}}(r)=\Big(1-\frac{2M}{r}\Big)\Big[\frac{l(l+1)}{r^2}+(1-s^2)\frac{2M}{r^3}\Big]\, .
\end{eqnarray}
Here, $t$ is the coordinate time, $M$ is the mass of the black hole, and $l$ is the angular momentum number from spherical harmonic decomposition. The index $s$ denotes the spin of the perturbed field, with $s=0$ for scalar, $s=1$ for electromagnetic, and $s=2$ for gravitational perturbations. The tortoise coordinate $x$ is related to the radial coordinate $r$ by
\begin{eqnarray}\label{tortoise_coordinate}
   \mathrm{d}x=\frac{\mathrm{d}r}{1-2M/r}\, ,\quad  x=r+2M\ln(r-2M)+ \text{constant}\, .
\end{eqnarray}
For convenience, the integration constant is chosen as zero. In the followings, we always use the unit $M=1$ and concentrate on the case $s=0$.

In~\cite{Nollert:1996rf}, the potential is approximated by the piecewise step function, while in~\cite{Daghigh:2020jyk}, the potential is approximated by the piecewise linear function. The approximation of the step function can be seen as a zero-order approximation, while the approximation of a piecewise linear function can be seen as a first-order approximation. So, naturally, what about second-order approximation? Another important reason for considering the parabolic approximation is that we can analytically solve the perturbation equation (see below).  

\begin{figure}[htbp]
	\centering
	\includegraphics[width=0.45\textwidth]{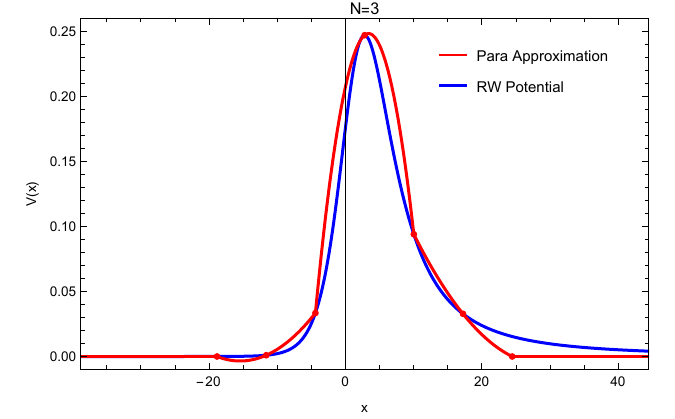}
 \hspace{0.5cm}
\includegraphics[width=0.45\textwidth]{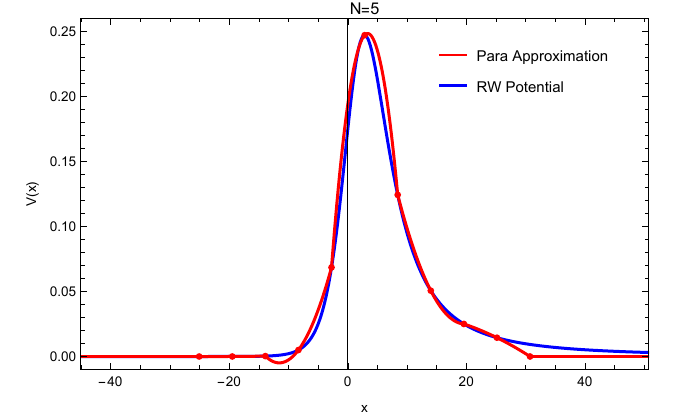}    
    \vspace{0.5cm}\\
    \includegraphics[width=0.45\textwidth]{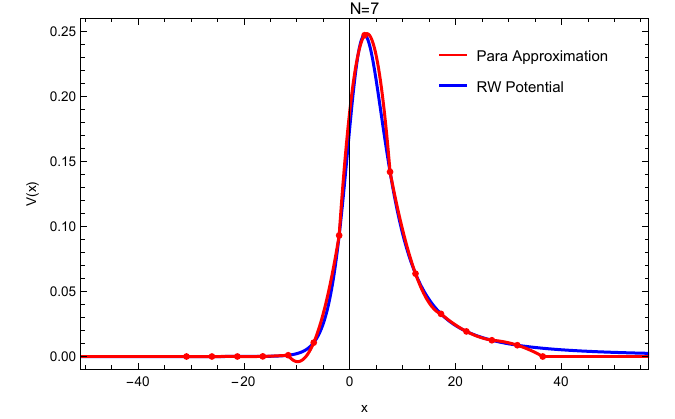}
 \hspace{0.5cm}
    \includegraphics[width=0.45\textwidth]{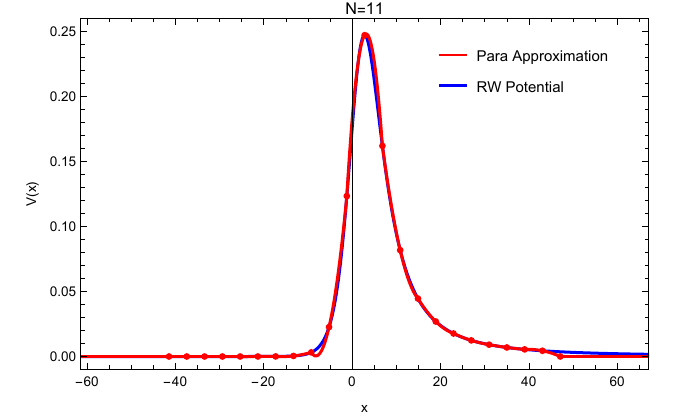} 
    \vspace{0.5cm}\\
    \includegraphics[width=0.45\textwidth]{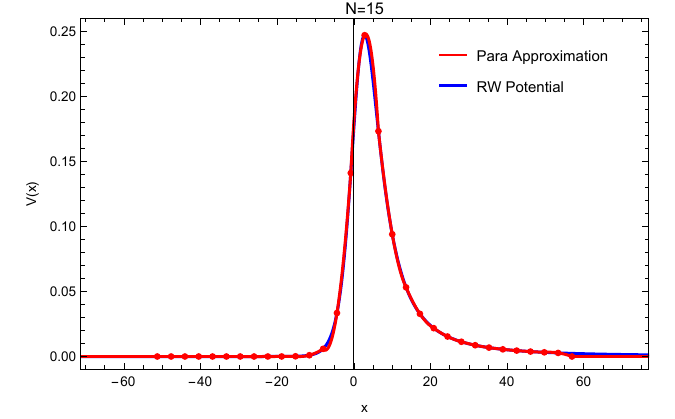}
 \hspace{0.5cm}
    \includegraphics[width=0.45\textwidth]{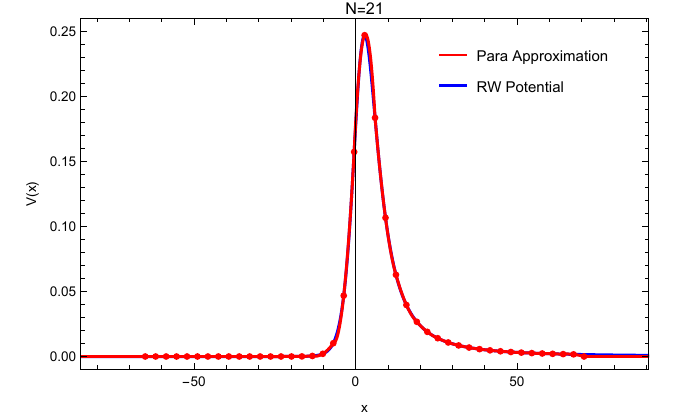}
	\caption{Comparison between the original Regge-Wheeler potential (blue curves) and the effective potential obtained via piecewise parabolic interpolation (red curves) for $l=2$ case with different segments $N$. Each panel shows both potentials as functions of the tortoise coordinate $x$, with increasing interpolation: from top-left to bottom-right, the number of interpolation segments is $N=3$, $5$, $7$, $11$, $15$, and $21$, respectively. As $N$ increases, the constructed potential converges toward the original R-W potential, illustrating the effectiveness of the piecewise parabolic approximation.}
	\label{fig_parabolic_approximation}
\end{figure}

\begin{figure}[htbp]
	\centering
	\includegraphics[width=0.45\textwidth]{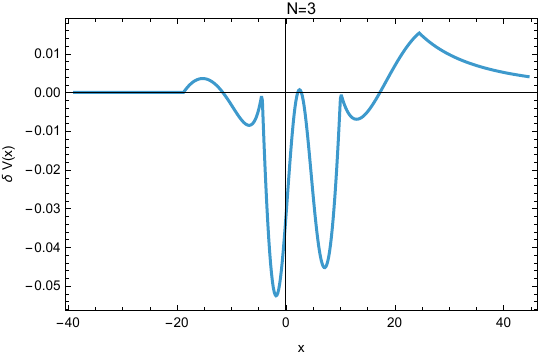}
 \hspace{0.5cm}
    \includegraphics[width=0.45\textwidth]{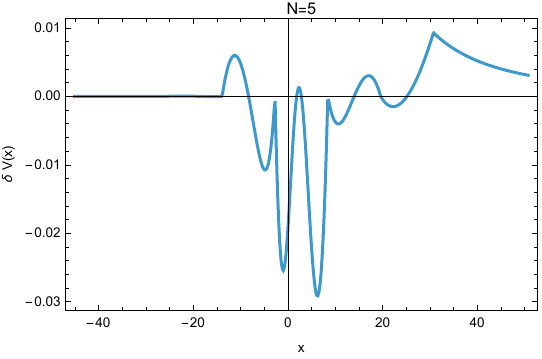}
    \vspace{0.5cm}\\
    \includegraphics[width=0.45\textwidth]{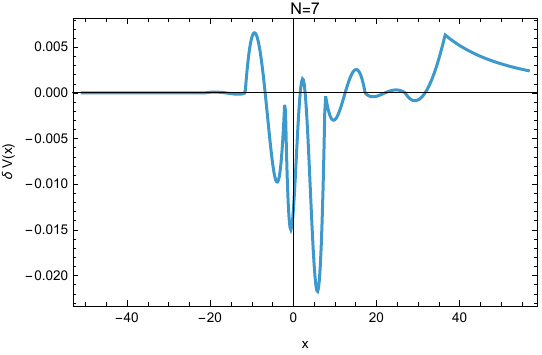}
 \hspace{0.5cm}
    \includegraphics[width=0.45\textwidth]{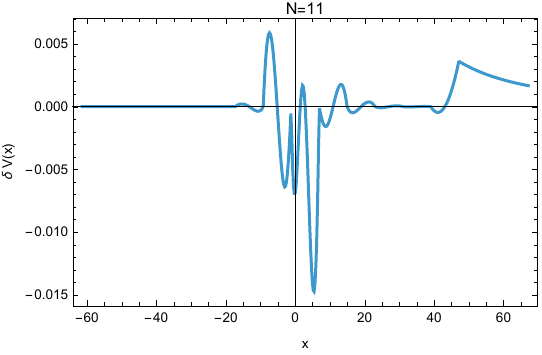}  
    \vspace{0.5cm}\\
    \includegraphics[width=0.45\textwidth]{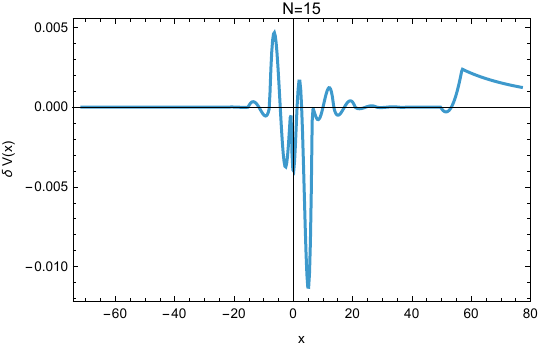}
 \hspace{0.5cm}
    \includegraphics[width=0.45\textwidth]{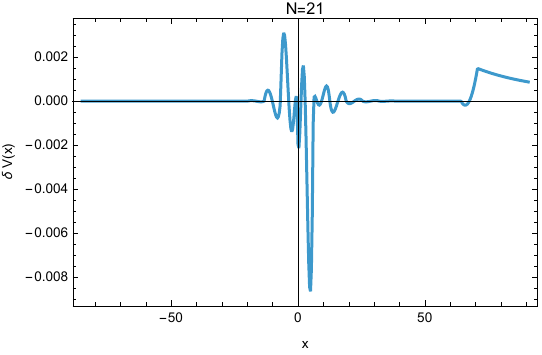}
	\caption{Difference between the original R-W potential and the effective potential obtained via piecewise parabolic approximation, where $N=3$, $5$, $7$, $11$, $15$, $21$. This plot shows $\delta V(x)=V_{\text{R-W}}(x)-V(x)$ as a function of the tortoise coordinate, illustrating the accuracy and local deviations introduced by the approximation procedure.}	\label{fig_parabolic_approximation_error}
\end{figure}

From now on, we will provide a specific strategy to achieve piecewise parabolic approximation of the effective potential. Here, we use the following parabolic approximation to the R-W potential in terms of the tortoise coordinate $x$ in each interval $[x_{2i-2},x_{2i}]$:
\begin{eqnarray}\label{parabolic_approximation}
    V(x)&=&V_{\text{R-W}}(x_{2k-2})\frac{(x-x_{2k-1})(x-x_{2k})}{(x_{2k-2}-x_{2k-1})(x_{2k-2}-x_{2k})}+V_{\text{R-W}}(x_{2k-1})\frac{(x-x_{2k-2})(x-x_{2k})}{(x_{2k-1}-x_{2k-2})(x_{2k-1}-x_{2k})}\nonumber\\
    &&+V_{\text{R-W}}(x_{2k})\frac{(x-x_{2k-2})(x-x_{2k-1})}{(x_{2k}-x_{2k-2})(x_{2k}-x_{2k-1})}\, ,\quad k=1,2,\cdots,N-1,N\, ,
\end{eqnarray}
where $N\ge1$ is the number of segments, and $V_{\text{R-W}}(x_{k})$ is the value of the R-W potential at $x_k$ for $k=1,\cdots,2N-1$. In addition, we choose $V_{\text{R-W}}(x_{0})=V_{\text{R-W}}(x_{2N})=0$. It is evident that we constructed effective potential is a $C^0$ function, but not a $C^1$ function. For simplicity, we require that the spacing between points be equal, and we denote such interval as $\delta x$. Subsequently, in order to achieve approximation on both sides of the original potential, we limit the number of parabolas to be an odd number. In this way, $x=x_{N}$ is always at the center of the point sequence $\{x_{k}\}_{k=0}^{2N}$, where its order number is $N+1$. In addition, $x_N$ is chosen to coincide with the peak of the Regge-Wheeler potential. More importantly, we need to ensure that the following two things occur simultaneously during the approximation process. The first thing is that as $N$ increases, the non-zero part of the approximated effective potential also increases. The second thing is that as $N$ increases, the interval between points decreases. It's not difficult to satisfy both of these things at the same time, we just need to appropriately define the relationship between the interval of points and $N$. Here, the relationship is introduced as
\begin{eqnarray}\label{relationship_interval_N}
    \delta x=\frac{10}{\ln (N+1)}\, .
\end{eqnarray}
Therefore, it can be found that the total length of the interval for the non-zero part of the approximated effective potential is $2N\delta x=20N/\ln(N+1)$. As a typical example, in Fig. \ref{fig_parabolic_approximation}, we present approximations of the R-W potential using varying numbers of parabolic segments, specifically for $N=3$, $5$, $7$, $11$, $15$, $21$ by using $l=2$. The most direct feeling is that as $N$ increases, the degree of approximation becomes better. The differences between these two effective potentials are described in Fig. \ref{fig_parabolic_approximation_error}. The differences will have the following characteristics: (1) the overall difference will be smaller; (2) The qualitative behavior near the peak of the effective potential is similar to the $\operatorname{sinc}$ function; (3) When $x$ approaches positive infinity, the behavior of the difference is $\sim x^{-2}$ (Due to our construction method, this is unavoidable.). The following two sections on spectrum instability, greybody factor stability, are both discussed based on our parabolic approximation strategy.

\section{The instability of the QNM spectra}\label{stability_QNMs}
In this section, we will compare with the QNM spectra from the R-W potential $V_{\text{R-W}}(r)$ and the parabolic approximation $V(x)$ to study the (in)stability of QNM spectra for the Schwarzschild black hole. Generally, on each interval $[x_{2k-2},x_{2k}]$, the approximated potential can be written as 
\begin{eqnarray}\label{potential_function_on_an_interval}
    V(x)=ax^2+bx+c\, ,\quad x\in[x_{2k-2},x_{2k}]\, ,
\end{eqnarray}
where $a$, $b$, $c$ can be read from Eq. (\ref{parabolic_approximation}). We assume to have harmonic time-dependence like $\Psi(t,x)\sim e^{-i\omega t}\Psi(x)$ (our convention), and modified R-W equation on the interval $[x_{2k-2},x_{2k}]$ becomes
\begin{eqnarray}\label{parabolic_potential_equation}
    \frac{\mathrm{d}^2\Psi_k(x)}{\mathrm{d}x^2}+\Big[\omega^2-(ax^2+bx+c)\Big]\Psi_k(x)=0\, ,
\end{eqnarray}
where the solution on interval $[x_{2k-2},x_{2k}]$ are denoted by $\Psi_k(x)$. One can get general solution of Eq. (\ref{parabolic_potential_equation}) on each interval $[x_{2k-2},x_{2k}]$, which means that 
\begin{eqnarray}\label{solution_parabolic_potential_equation}
    \Psi_k(x)=C_{1, k} D_{\nu}(tx+s)+C_{2, k} D_{-\nu-1}(i(tx+s))\, ,\quad t=(4a)^{1/4}\, ,\quad s=\frac{b}{2a}(4a)^{1/4}\, ,\quad x\in[x_{2k-2},x_{2k}]\, ,
\end{eqnarray}
where $D_\nu$ is called the parabolic cylinder function, and $C_{1, k}$, $C_{2, k}$ are constants, $k=1,\cdots,N$. The detail can be found in Appendix \ref{solution_parabolic_potential}. According to the boundary conditions of the QNMs, for $x<x_0$, the solution is $\Psi(x)=Ae^{-i\omega x}$ while for $x>x_{2N}$, the solution is $\Psi(x)=Be^{i\omega x}$. Since our constructed potentials $V(x)$ are $C^0$ functions, the solution $\Psi(x)$ is at least a $C^1$ function. Therefore, the QNM spectra are determined by imposing the following conditions (junction conditions)
\begin{eqnarray}\label{QNM_conditions}
    \Psi(x)\big|_{x\to x_{2k}^{-}}&=&\Psi(x)\big|_{x\to x_{2k}^{+}}\nonumber\\
    \Psi^{\prime}(x)\big|_{x\to x_{2k}^{-}}&=&\Psi^{\prime}(x)\big|_{x\to x_{2k}^{+}}\, ,
\end{eqnarray}
at the points $x_{2k}$, in which $k=0,1,\cdots,N-1,N$. It is worth mentioning that when the potential is discontinuous, the junction conditions will be different~\cite{Shen:2022xdp,Li:2024npg,Li:2025ljb}. There are $2N+2$ conditions in Eqs. (\ref{QNM_conditions}), and it is not difficult to find that these conditions are linear for coefficients $A$, $C_{1,k}$, $C_{2,k}$, $B$, $k=1,\cdots,N$. Here, the prime stands for the derivative with respect to $x$. The corresponding coefficient matrix is denoted by $\mathbf{M}(\omega)$. The QNM spectra are solved by the condition of vanishment of the determinant of $\mathbf{M}(\omega)$, i.e.,
\begin{eqnarray}\label{det_M}
    \det(\mathbf{M}(\omega))=0\, .
\end{eqnarray}
In Appendix \ref{construction_M} (taking $N=3$ as an example), we give the detail of the coefficient matrix $\mathbf{M}(\omega)$. Note the size of $\mathbf{M}$ is $(2N+2)\times(2N+2)$.

Unfortunately, the elements of matrix $\mathbf{M}(\omega)$ are non-linear functions in terms of $\omega$. The determinant function $\det(\mathbf{M}(\omega))$ is also highly non-linear. This is very unfavorable for us to solve the equation (\ref{det_M}). Since the approximation potential is segmented smooth, multi-domain spectral collocation method~\cite{PanossoMacedo:2024pox} is a reliable and effective method to solve the above QNM problems currently encountered. From its name may be derived its meaning, one uses spectral collocation method on each intervals $[x_{2k-2},x_{2k}]$, where the Chebyshev-Lobatto grids~\cite{Trefethen_spectral_method_book} 
\begin{eqnarray}\label{CL_grids}
    x^{(k)}_{i}=\frac{x_{2k-2}+x_{2k+2}}{2}+\frac{x_{2k+2}-x_{2k-2}}{2}\cos\Big(\frac{i\pi}{N_{\text{g}}}\Big)\, ,\quad i=0,\cdots,N_{\text{g}}
\end{eqnarray}
are used on each intervals. For simplicity, the resolutions of each interval remain the same. Accordingly, the solutions $\{\Psi_1(x),\cdots,\Psi_N(x)\}$ are flattened into a vector denoted by $\mathbf{\Psi}$ whose dimension is $N\times(N_\text{g}+1)$. Within the interval (excluding boundaries), we obtain the following conditions,
\begin{eqnarray}\label{inner_conditions}
    \Big[\partial_{x}^2+(\omega^2-V)\Big]\Psi_{k}(x_i^{(k)})=0\, ,\quad i=1,\cdots,N_{\text{g}}-1\, .
\end{eqnarray}
The above equations give $N\times(N_\text{g}-1)$ conditions for $\mathbf{\Psi}$. Because when $x<x_0$, $\Psi(x)=Ae^{-i\omega x}$ and when $x>x_{2N}$, $\Psi(x)=Be^{i\omega x}$, by using Eqs. (\ref{QNM_conditions}), one obtains that
\begin{eqnarray}\label{left_condition}
    \Big(i\omega\Psi_1+\partial_x\Psi_1\Big)(x_{N_{\text{g}}}^{(1)})=0\, ,
\end{eqnarray}
and
\begin{eqnarray}\label{right_condition}
    \Big(-i\omega\Psi_N+\partial_x\Psi_N\Big)(x_{0}^{(N)})=0\, .
\end{eqnarray}
More importantly, at the interfaces among adjacent intervals, we have junction conditions (\ref{QNM_conditions}). That is to say that another $2(N-1)$ conditions will be found, i.e.,
\begin{eqnarray}\label{interface_condition}
    \Psi_k(x_{N_{\text{g}}}^{(k)})-\Psi_{k+1}(x_0^{(k+1)})=0\, ,\quad (\partial_x\Psi_k)(x_{N_{\text{g}}}^{(k)})-(\partial_x\Psi_{k+1})(x_0^{(k+1)})=0\, ,\quad k=1,\cdots,N-1\, , 
\end{eqnarray}
Finally, replacing the derivatives with the differential matrices on the corresponding interval, a homogeneous system of equations about $\mathbf{\Psi}$ will be obtained, which is expressed as $\mathbf{T}(\omega)\mathbf{\Psi}=0$. Therefore, the matrix $\mathbf{T}(\omega)$ contains both information about differential equations and information about boundary conditions or interface junction conditions.  Subsequently, the QNM spectra are attained at vanishing determinant
\begin{eqnarray}\label{det_T_0}
    \det(\mathbf{T}(\omega))=0\, .
\end{eqnarray}
where the notation of such coefficient matrix is denoted by $\mathbf{T}(\omega)$ whose dimension is $[N(N_\text{g}+1)\times N(N_\text{g}+1)]$. Generally speaking, $N_{\text{g}}$ is much larger than $N$, which results in the dimension of the matrix $\mathbf{T}(\omega)$ being much larger than that of the matrix $\mathbf{M}(\omega)$. But conversely, the structure of matrix elements for $\mathbf{T}(\omega)$ becomes much simpler. The matrix elements of $\mathbf{T}(\omega)$ are at most quadratic polynomials of $\omega$. By the way, the objects of action for matrices $\mathbf{M}$ and $\mathbf{T}$ are completely different. The former is the coefficient of the parabolic cylinder functions (see Appendix \ref{construction_M}), while the latter is the value of the solution function at the grid points. In~\cite{Jansen:2017oag}, an effective algorithm is proposed. According to the approach given from~\cite{Jansen:2017oag,Li:2024npg,Li:2025ljb},  we rewrite the matrix equation coming from Eqs. (\ref{inner_conditions})-(\ref{interface_condition}) in the form
\begin{eqnarray}\label{matrix_T}
    \mathbf{T}(\omega)\mathbf{\Psi}=\Big(\mathbf{T}_0+\mathbf{T}_1\omega+\mathbf{T}_2\omega^2\Big)\mathbf{\Psi}=0\, .
\end{eqnarray}
It is straightforward to show that Eq. (\ref{matrix_T}) can be formally written as
\begin{eqnarray}\label{general_eigenvalue_problem}
    \Big(\tilde{\mathbf{T}}_0-\omega\tilde{\mathbf{T}}_1\Big)\tilde{\mathbf{\Psi}}=0\, ,
\end{eqnarray}
in which
\begin{eqnarray}
    \tilde{\mathbf{T}}_0=
    \begin{bmatrix}
        \mathbf{T}_0 & \mathbf{T}_1 \\
        \mathbf{0} & \mathbf{I}
    \end{bmatrix}\, ,\quad \tilde{\mathbf{T}}_1=
    \begin{bmatrix}
        \mathbf{0} & -\mathbf{T}_2\\
        \mathbf{I} & \mathbf{0}
    \end{bmatrix}\, ,\quad \tilde{\mathbf{\Psi}}=
    \begin{bmatrix}
        \mathbf{\Psi}\\
        \omega\mathbf{\Psi}
    \end{bmatrix}\, ,
\end{eqnarray}
and $\tilde{\mathbf{T}}_0$, $\tilde{\mathbf{T}}_1$ are both $2N(N_{\text{g}}+1)\times2N(N_{\text{g}}+1)$ matrices. Eq. (\ref{general_eigenvalue_problem}) is a general eigenvalue problem associated with $\omega$, which can be solved using \textit{Mathematica}, like \textit{Eigenvalue[{$\{\cdot,\cdot\}$}]}.

\begin{figure}[htbp]
	\centering
	\includegraphics[width=0.45\textwidth]{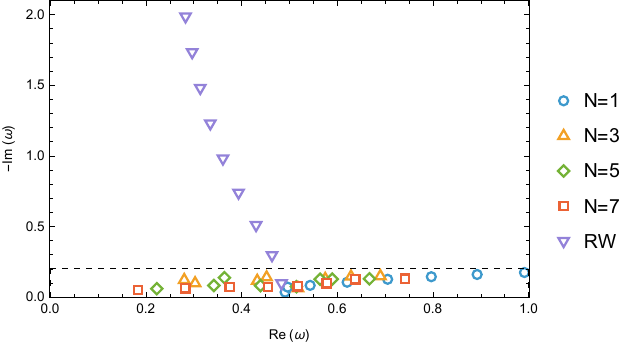}
    \hspace{0.5cm}
    \includegraphics[width=0.45\textwidth]{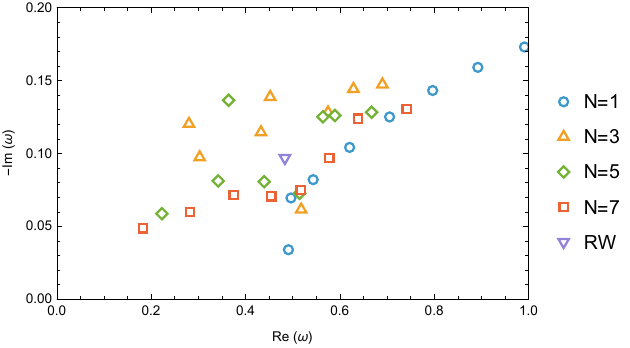}    
	\caption{Left: For $l=2$, QNM spectra provided in Tab. \ref{QNM_spectra}, are plotted, where the value of the horizontal dashed line is $0.2$. Right: Zoom-in of the left panel for the region with $-\text{Im}(\omega)\in[0,0.2]$.}
	\label{QNMs_Plot}
\end{figure}

In Tab. \ref{QNM_spectra}, for $l=2$, we show the QNM spectra of the approximated potential with $N=1$, $N=3$, $N=5$ and $N=7$ comparing with the QNM spectra of the scalar perturbation for the Schwarzschild black hole, where the overtone number has reached $n=7$. For comparison, we also include the first eight QNM spectra for $V_{l=2}^{\text{scalar}}$ using the continued fraction method~\cite{Leaver:1985ax}. The spectra for $V_{l=2}^{\text{scalar}}$ discovered are consistent with other techniques, like the pseudo-spectral method within the hyperboloidal framework~\cite{Jaramillo:2020tuu,PanossoMacedo:2023qzp,PanossoMacedo:2024nkw}. In Fig. \ref{QNMs_Plot}, we plot the data of Tab. \ref{QNM_spectra}. 

As $N$ increases, the effective potential generated by the piecewise parabolic approximation asymptotically converges toward the R-W potential, yet the corresponding QNM spectrum manifests an other behavior. Specifically, the results reveal a strikingly different feature of the QNM spectra associated with the parabolic approximation potential as compared to those of the original R-W case: As shown in Fig. \ref{QNMs_Plot}, for any parabolic approximated potential with different $N$, the resulting QNM spectra exhibit imaginary parts that decrease slowly (i.e., become more negative) with increasing overtone number $n$, remaining close to zero. These behaviors contrast sharply to the case of original R-W potential, where the imaginary parts decrease rapidly. The above phenomenon is the instability of the spectra. Such overtones with slowly decreasing (weakly damped) imaginary parts are referred to as long-lived modes. This behavior is also observed in cases with discontinuous effective potentials~\cite{Qian:2020cnz} or piecewise linear approximation potential~\cite{Daghigh:2020jyk}. It is worth mentioning that the piecewise linear approximation used in Ref. \cite{Daghigh:2020jyk} uses $\delta x$ fixed as a constant (do not dependent on $N$), while the $\delta x$ we use is not a constant, but a number that depends on $N$ [see Eq. (\ref{relationship_interval_N})]. One salient advantage of our present approach lies in the progressive diminishment of discrepancies in the vicinity of the effective potential's maximum as $N$ is augmented making the approximated effective potential closer to the original R-W potential.

\begin{table}[ht]
 \renewcommand{\arraystretch}{1.5}
\begin{tabular}{lccccc}
 \hline
 
$n$ ~~~& $N=1$~~~& $N=3$ ~~~& $N=5$ ~~~& $N=7$ ~~~& $V_{l=2}^{\text{scalar}}$\cr

 \hline
 0 & $0.4912-\imath~0.0339$ ~~~& $0.5182-\imath~0.0614$ ~~~& $0.2232-\imath~0.0586$ ~~~& $0.1833-\imath~0.0485$ ~~~& $0.4836-\imath~0.0968$\\
 1 & $0.4966-\imath~0.0694$ ~~~& $0.3031-\imath~0.0973$ ~~~& $0.5149-\imath~0.0726$ ~~~& $0.2829-\imath~0.0598$ ~~~& $0.4639-\imath~0.2956$\\
 2 & $0.5436-\imath~0.0819$ ~~~& $0.4330-\imath~0.1145$ ~~~& $0.4398-\imath~0.0805$ ~~~& $0.4550-\imath~0.0706$ ~~~& $0.4305-\imath~0.5086$\\
  3 & $0.6209-\imath~0.1041$ ~~~& $0.2803-\imath~0.1202$ ~~~& $0.3423-\imath~0.0811$ ~~~& $0.3752-\imath~0.0716$ ~~~& $0.3939-\imath~0.7381$ \\
 4 & $0.7055-\imath~0.1250$ ~~~& $0.5751-\imath~0.1281$ ~~~& $0.5644-\imath~0.1251$ ~~~& $0.5172-\imath~0.0749$ ~~~& $0.3613-\imath~0.9799$\\
 5 & $0.7968-\imath~0.1431$ ~~~& $0.4525-\imath~0.1387$ ~~~& $0.5898-\imath~0.1259$ ~~~& $0.5781-\imath~0.0967$ ~~~& $0.3349-\imath~1.2284$ \\
 6 & $0.8926-\imath~0.1590$ ~~~& $0.6289-\imath~0.1442$ ~~~& $0.6674-\imath~0.1282$ ~~~& $0.6382-\imath~0.1238$ ~~~& $0.3139-\imath~1.4800$\\
 7 & $0.9915-\imath~0.1729$ ~~~& $0.6904-\imath~0.1473$ ~~~& $0.3644-\imath~0.1364$ ~~~& $0.7415-\imath~0.1304$ ~~~& $0.2970-\imath~1.7328$\\
 \hline
\end{tabular}
\caption{For $l=2$, the QNM spectra of the approximation potential with $N=1$, $N=3$, $N=5$ and $N=7$ compare with the QNM spectra of the scalar perturbation. The results of the first four columns are all obtained from the resolution $N_{\text{g}}=60$. All results are rounded to $4$ decimal places. The sorting of overtones is still based on the absolute of the imaginary part.}
\label{QNM_spectra}
\end{table}

\begin{figure}[htbp]
	\centering
	\includegraphics[width=0.45\textwidth]{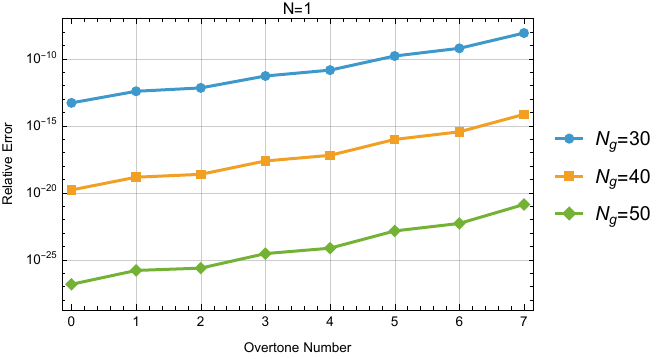}
 \hspace{0.5cm}
    \includegraphics[width=0.45\textwidth]{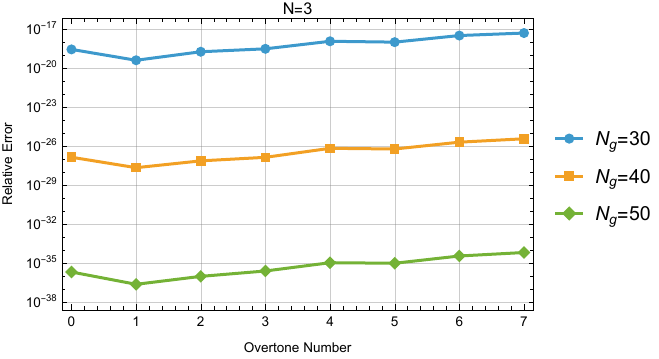}\\   
    \vspace{0.5cm}
    \includegraphics[width=0.45\textwidth]{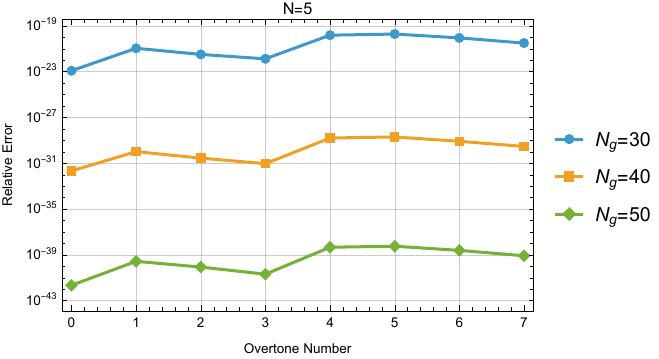}
 \hspace{0.5cm}
    \includegraphics[width=0.45\textwidth]{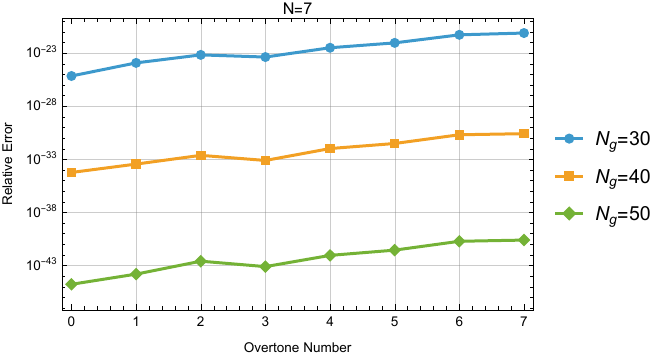} 
	\caption{For $l=2$, the relative errors are shown, where the horizontal axis represents overtone numbers, and the resolution $N_{\text{g}}=60$ is the benchmark. Different colored lines correspond to different resolutions, where blue corresponds to $N_{\text{g}}=30$, green corresponds to $N_{\text{g}}=40$, and yellow corresponds to $N_{\text{g}}=50$.}
	\label{relative_errors}
\end{figure}

Finally, we close this section with checking the accuracy of the results. In the previous content, we obtained that the condition for QNM spectra is given by $\det(\mathbf{M}(\omega))=0$ [see Eq. (\ref{det_M})]. Notwithstanding the intractability of this condition for direct computation of QNM spectra, it remains rigorously applicable for validation. Operationally, this entails substituting QNM spectra derived via the multi-domain spectral collocation method into Eq. (\ref{det_M}) and subsequently computing the modulus of the determinant, i.e., $|\det(\mathbf{M}(\omega))|$. The moduli of the determinants are shown in Tab. \ref{QNM_spectra_det_M} with $N=1$, $N=3$, $N=5$ and $N=7$. We can see that all values (The maximum is about $10^{-32}$.) are very small, ensuring the accuracy of the QNM spectrum. Another way to ensure the accuracy of QNM spectra is to use different resolutions to calculate relative errors, where we use the results with a resolution of $N_{\text{g}}=60$ as the benchmark. In Fig. \ref{relative_errors}, we depict the relative errors. Each panels have different numbers of parabolic segments. As the number of overtones increases, the relative error generally increases gradually. The results in Fig. \ref{relative_errors} also provide evidences of computational accuracy.


\begin{table}[ht]
 \renewcommand{\arraystretch}{1.5}
\begin{tabular}{lcccc}
 \hline
 
$n$ ~~~& $N=1$~~~& $N=3$ ~~~& $N=5$ ~~~& $N=7$\cr

\hline
0 & $2.0885\times10^{-34}$ ~~~& $3.4472\times10^{-49}$ ~~~& $3.0607\times10^{-57}$ ~~~& $3.1304\times10^{-76}$\\
1 & $2.8532\times10^{-33}$ ~~~& $2.2467\times10^{-47}$ ~~~& $1.0822\times10^{-76}$ ~~~& $3.4341\times10^{-94}$\\
2 & $1.1267\times10^{-33}$ ~~~& $8.0995\times10^{-50}$ ~~~& $1.7139\times10^{-72}$ ~~~& $1.3004\times10^{-123}$\\
3 & $3.7305\times10^{-33}$ ~~~& $4.3749\times10^{-46}$ ~~~& $6.9155\times10^{-66}$ ~~~& $3.6952\times10^{-117}$\\
4 & $1.1627\times10^{-33}$ ~~~& $1.1760\times10^{-51}$ ~~~& $1.2391\times10^{-89}$ ~~~& $6.3620\times10^{-115}$\\
5 & $1.8567\times10^{-33}$ ~~~& $1.7212\times10^{-49}$ ~~~& $2.2582\times10^{-92}$ ~~~& $6.3573\times10^{-89}$\\
6 & $2.5120\times10^{-34}$ ~~~& $1.0271\times10^{-52}$ ~~~& $2.8116\times10^{-101}$ ~~~& $9.7495\times10^{-55}$\\
7 & $1.6802\times10^{-34}$ ~~~& $9.8135\times10^{-54}$ ~~~& $9.0499\times10^{-72}$ ~~~& $4.1363\times10^{-32}$\\
 \hline
\end{tabular}
\caption{For $l=2$, the moduli of the determinants, $|\det(\mathbf{M}(\omega))|$, are shown, where $\omega$ is choosen from the first four columns of Tab. \ref{QNM_spectra}.}
\label{QNM_spectra_det_M}
\end{table}

\section{The stability of the greybody factors}\label{sec: GFs}
Greybody factors demonstrate enhanced robustness in response to minor perturbations of the underlying geometric configurations relative to their QNM counterparts, and there are many studies on the stability of greybody factors~\cite{Oshita:2024fzf,Wu:2024ldo,Rosato:2024arw,Ianniccari:2024ysv}. Such structural invariability endows greybody factors with unique diagnostic utility for interrogating the spacetime architecture. In Refs. \cite{Oshita:2024fzf,Wu:2024ldo,Rosato:2024arw,Ianniccari:2024ysv}, a bump is added to the original effective potential, and the greybody factors are stable for such approach. How will the non-smooth correction of the effective potential affect the greybody factor, as the above mentioned method of adding small bumps smoothly changes the effective potential? Here, in this section, we consider the stability of greybody factors for the approximation potential (\ref{parabolic_approximation}), where such modification can be viewed as a non-smooth change to the potential. 

First of all, we give a brief review on the greybody factor. The greybody factor quantifies the absorptive nature of a black hole, and it is also determined by the geometry of black holes like the QNM spectra. The black hole greybody factor is the transmission coefficient of a scattering problem identified by the following boundary condition
\begin{eqnarray}\label{boundary_condition_GF}
    \Psi(x)=\left\{\begin{array}{l}
    e^{-i\omega x}\, ,\quad x \to-\infty \\
    A^{\text {in }}(\omega) e^{-i\omega x}+A^{\text {out }}(\omega) e^{+i\omega x}\, ,\quad x\to+\infty
\end{array}\right.\, .
\end{eqnarray}
Here, the frequency $\omega\in\mathbb{R}$ for the study on the greybody factors.  $\Psi(x)$ is consist of purely ingoing waves at the event horizon and a mixture of ingoing and outgoing waves at spatial infinity. When $\omega\in\mathbb{C}$, the QNM spectra are defined as $A^{\text{in}}(\omega)=0$. For the scattering problem with an effective potential, we can define the reflection coefficient and transmission coefficient as follows
\begin{eqnarray}\label{reflectivity_transmissivity}
    R_l(\omega)=\Bigg|\frac{A^{\text{out}}(\omega)}{A^{\text {in}}(\omega)}\Bigg|^2, \quad \Gamma_l(\omega)=\Bigg|\frac{1}{A^{\text {in}}(\omega)}\Bigg|^2\, ,
\end{eqnarray}
in which $\Gamma_l(\omega)$ is the greybody factor and energy conservation enforces $R_l(\omega)+\Gamma_l(\omega)= 1$. With $\omega\in\mathbb{R}$, we use traditional $4$-order Runge-Kutta method to solve the R-W equation (\ref{Regge_Wheeler_equation}) in the frequency-domain with the boundary condition (\ref{boundary_condition_GF}) to get the greybody factor $\Gamma_{l}^{\text{R-W}}(\omega)$, which will be used for comparison.

As stated in Sec. \ref{stability_QNMs}, Eq. (\ref{parabolic_potential_equation}) admits an analytical solution when the potential function takes the form of Eq. (\ref{parabolic_approximation}). We can analytically calculate the greybody factor by imposing the boundary conditions Eq. (\ref{boundary_condition_GF}). Given such boundary condition, $\Psi(x)$ takes the following analytical form in different intervals,

\begin{eqnarray}\label{boundary_condition}
    \Psi(x)=\left\{\begin{array}{l}
    \Psi_{\text{left}}(x)=e^{-i\omega x}\, ,\quad x \leq x_{0} \\
    \Psi_{k}(x)=C_{1,k}(\omega)D_{\nu}(tx+s)+C_{2,k}(\omega) D_{-\nu-1}(i(tx+s))\, ,\quad x\in[x_{2k-2},x_{2k}]\\
    \Psi_{\text{right}}(x)=A^{\text {in }}(\omega) e^{-i\omega x}+A^{\text {out }}(\omega) e^{+i\omega x}\, ,\quad x\geq x_{2N}
    \end{array}\right.\, .
\end{eqnarray}
Here, $k$ denotes the index of each interval, and $k=1,2,\cdots,N-1,N$. To compute the greybody factor, we start from $\Psi_{\text{left}}(x)$ and apply the matching conditions at each interface $x_j$. At each $x_j$, two matching boundary conditions allow us to iteratively determine $C_{1,k}(\omega)$ and $C_{2,k}(\omega)$. This matching procedure proceeds iteratively across each interval and concludes at $x_{2N}$, where the coefficients $A^{\text{in}}(\omega)$ and $A^{\text{out}}(\omega)$ are determined. The greybody factor $\Gamma_l^{\text{Para}}(\omega)$ is then computed using Eq.~(\ref{reflectivity_transmissivity}). For the piecewise parabolic potential as our considered model, the solutions $A^{\text{in}}(\omega)$ and $A^{\text{out}}(\omega)$ about $\omega$ have a matrix representation (\ref{Ain_Aout}), which can be solved analytically, and for this time $\omega\in\mathbb{C}$. In addition, according to the definition of QNMs, we can also use some root-finding process to obtain the QNM spectra for the function $A^{\text{in}}(\omega)$. Since we have solved the QNM spectra by using the multi-domain spectral collocation method, $A^{\text{in}}(\omega)=0$ is able to be used to check our results. One can find more details in the Appendix \ref{app: greybody factor}. 

\begin{figure}[htbp]
	\centering
    \includegraphics[width=0.45\textwidth]{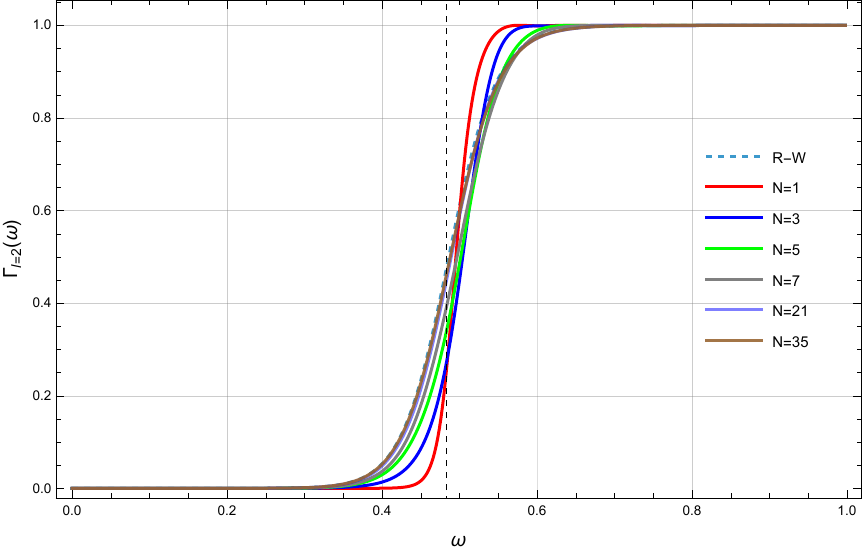}
    \hspace{0.5cm}
    \includegraphics[width=0.45\textwidth]{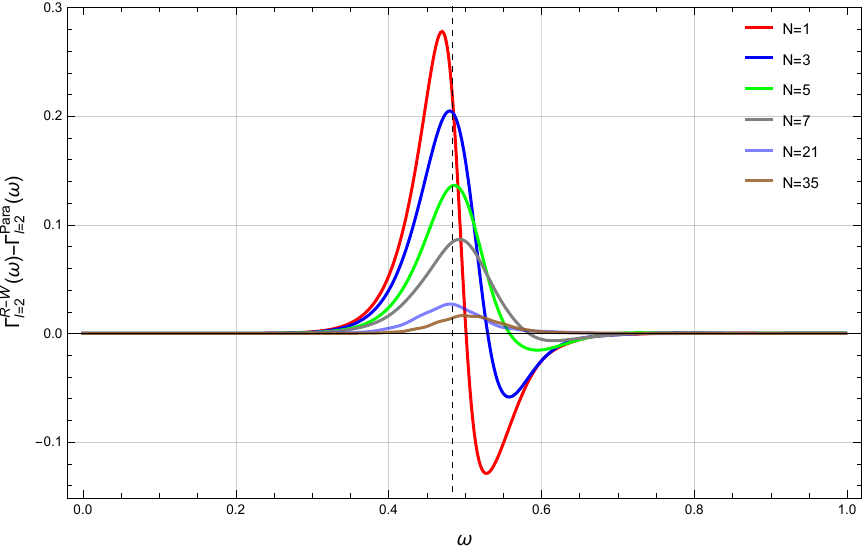}\\
    \vspace{0.5cm}
    \includegraphics[width=0.45\textwidth]{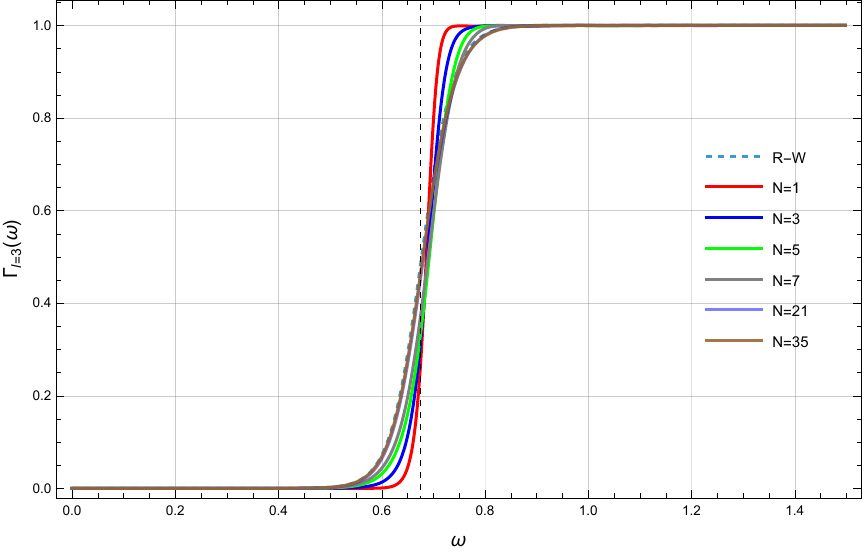}
    \hspace{0.5cm}
    \includegraphics[width=0.45\textwidth]{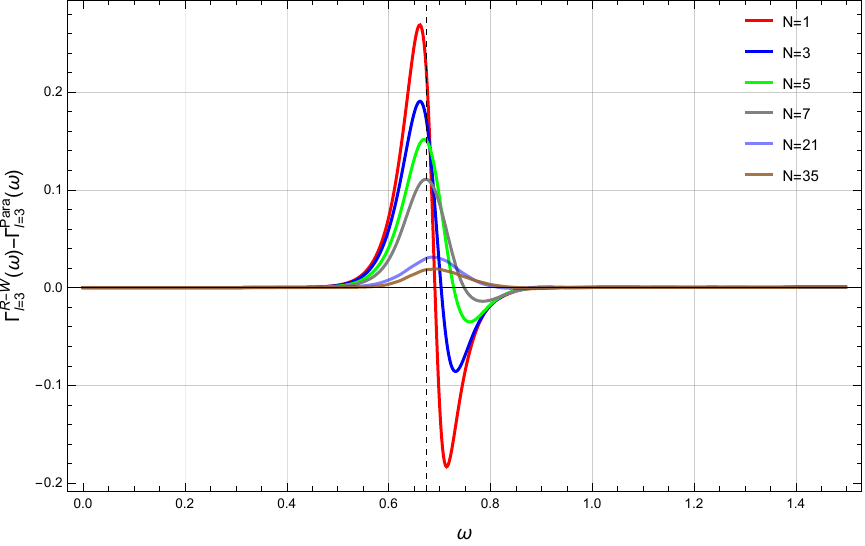}  
    \caption{Left: Comparison of the greybody factors between the R-W potential and the piecewise parabolic approximation potential, where the top corresponds to the case $l=2$ and the bottom corresponds to the case $l=3$. Right: Greybody factors' differences between the R-W potential and the piecewise parabolic approximation potential, where the top corresponds to the case $l=2$ and the bottom corresponds to the case $l=3$.}
	\label{GFs_Plot}
\end{figure}

\begin{figure}[htbp]
	\centering
    \includegraphics[width=0.45\textwidth]{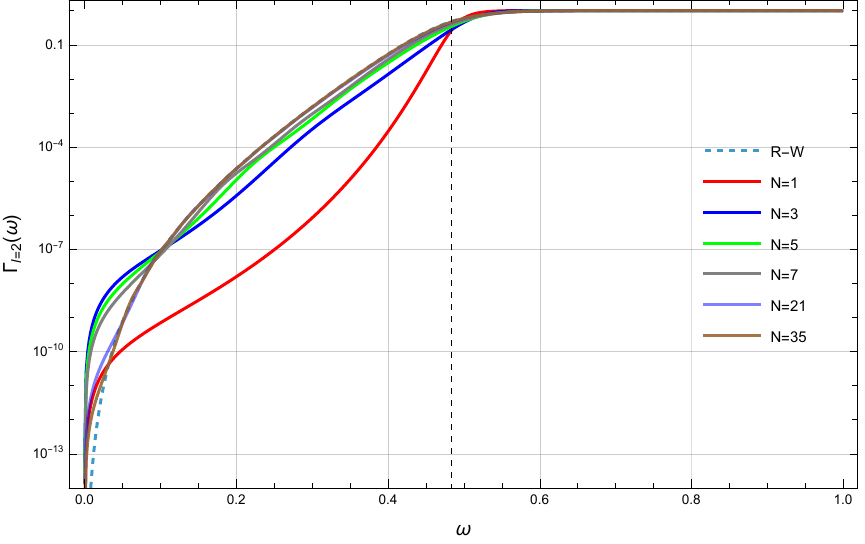}
    \hspace{0.5cm}
    \includegraphics[width=0.45\textwidth]{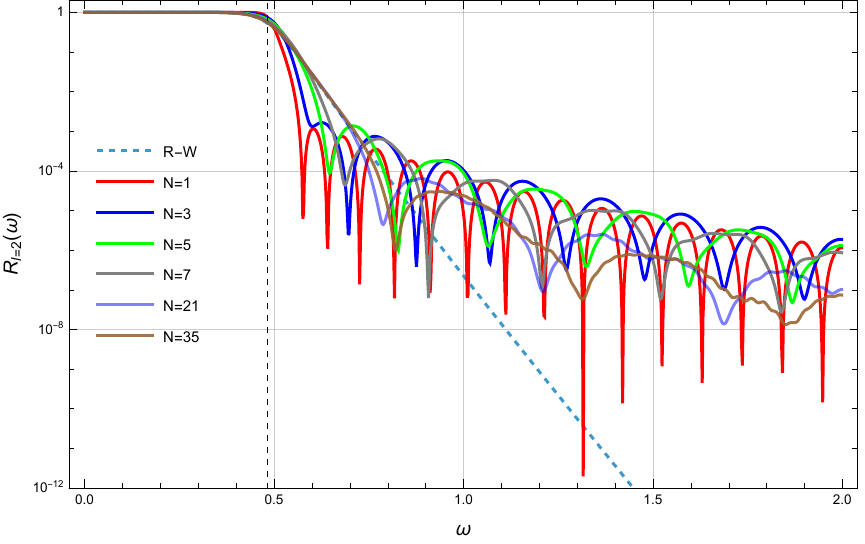}\\
    \vspace{0.5cm}
    \includegraphics[width=0.45\textwidth]{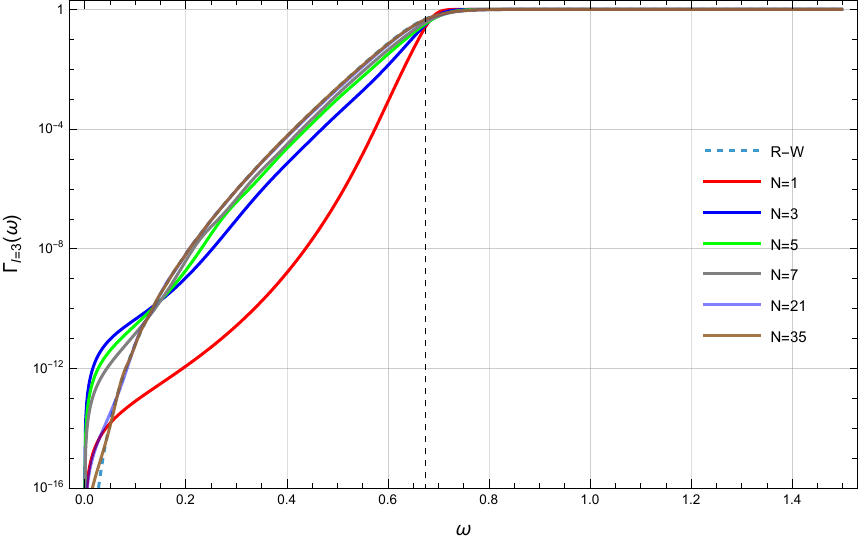}
    \hspace{0.5cm}
    \includegraphics[width=0.45\textwidth]{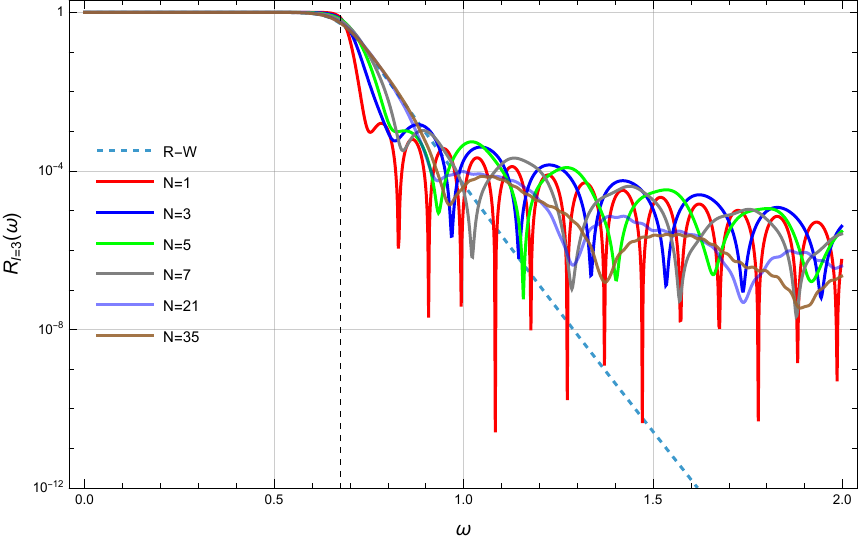}
    \caption{Top: Logplots for Comparisons of the greybody factors the reflectivities between the R-W potential and the piecewise parabolic approximation potential, with the case $l=2$,  where $N=1$, $3$, $5$, $7$, $21$, and $35$. Bottom: Logplots for Comparisons of the greybody factors and the reflectivities between the R-W potential and the piecewise parabolic approximation potential, with the case $l=3$, where $N=1$, $3$, $5$, $7$, $21$, and $35$.}
    \label{GFs_Reflection_Log}
\end{figure}

For different numbers of segments, we provide the greybody factors by using Eq. (\ref{Ain_Aout}) in Fig. \ref{GFs_Plot}, where the right panels of Fig. \ref{GFs_Plot} show the differences between the greybody factor of R-W potential and the greybody factors of piecewise parabolic potentials. As shown in Fig. \ref{GFs_Plot}, the results tell us that as the piecewise parabolic potential approaches the R-W potential, $\Gamma_l^{\text{Para}}(\omega)$ also approaches $\Gamma_{l}^{\text{R-W}}(\omega)$. Therefore, such results tell us that the greybody factor is stable under the non-smooth changes ($C^1$ discontinuity) of the potential. In addition, the main differences are reflected in the vicinity of the real part of the fundamental mode $\omega_0^{\text{R-W}}$, and the vertical dashed line in Fig. \ref{GFs_Plot} corresponds to $\omega_0^{\text{R-W}}$ (also see Fig. \ref{GFs_Reflection_Log}).
 
Furthermore, we focus on analyzing the behavior of the greybody factors in the low and high frequency regimes. So in Fig. \ref{GFs_Reflection_Log}, we present logarithmic plots of the greybody factor and the reflection coefficient as a function of the frequency $\omega$. As shown in the left panel of Fig. \ref{GFs_Reflection_Log}, in the low-frequency regime, the greybody factor converges well to the case of the original R-W potential as the number of parabolic segments $N$ increases. In the high-frequency regime, as evident in the right panel of Fig. \ref{GFs_Reflection_Log}, the reflection coefficient exhibits oscillation that strikingly differs from the result of R-W potential, where the oscillation period monotonically increasing as the number of segments increases.

\begin{figure}[htbp]
	\centering
    \includegraphics[width=0.45\textwidth]{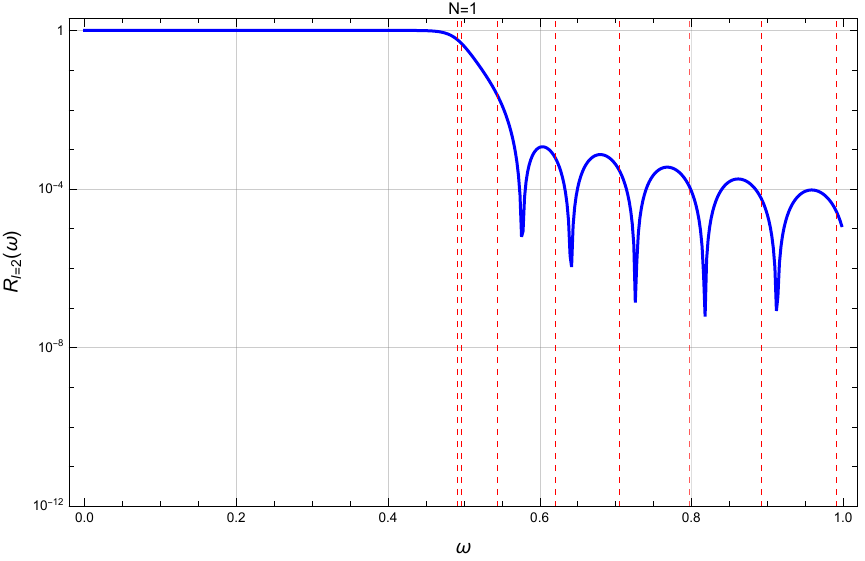}
    \hspace{0.5cm}
    \includegraphics[width=0.45\textwidth]{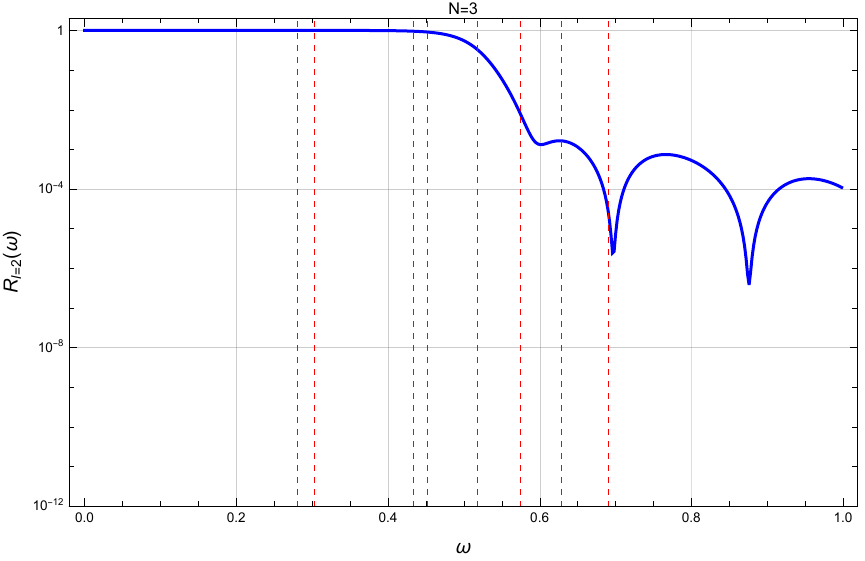}\\
    \vspace{0.5cm}
    \includegraphics[width=0.45\textwidth]{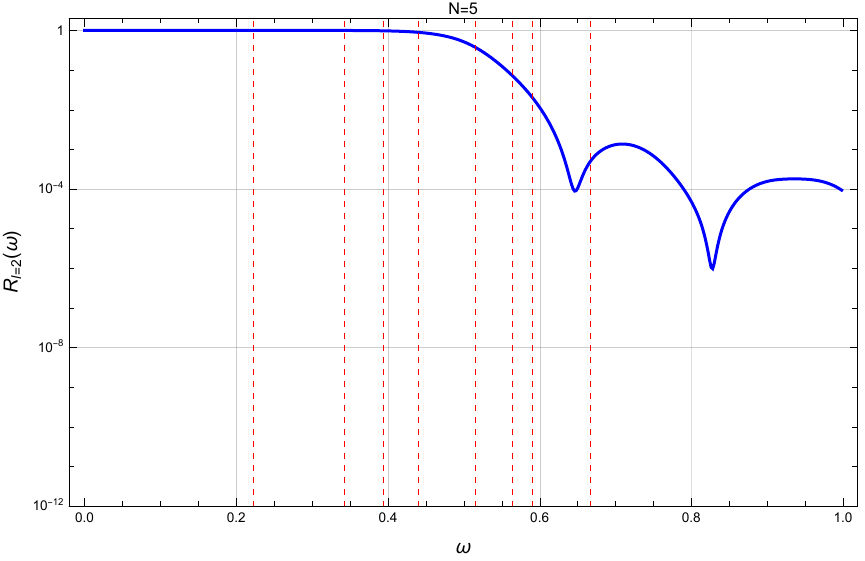}
    \hspace{0.5cm}
    \includegraphics[width=0.45\textwidth]{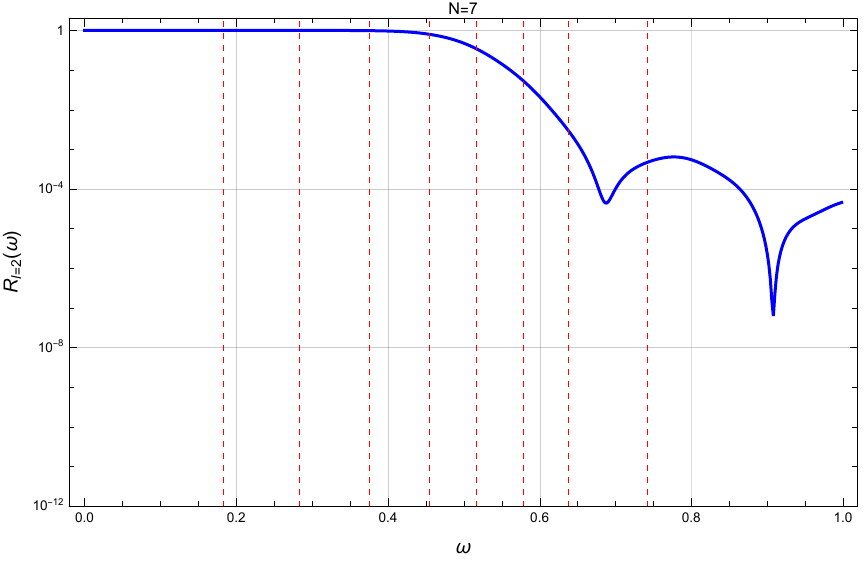}
    \caption{For $l=2$, comparisons between the real part of the QNM spectra (from Tab. \ref{QNM_spectra}) and the resonances displayed by the reflectivity of the piecewise parabolic approximated potential with $N=1$, $3$, $5$, and $7$.}
    \label{R_Log_QNMs}
\end{figure}

It is worth noting that such high-frequency resonance features in the reflection coefficient also appear in certain ultra-compact horizonless objects and are believed to be directly related to echoes~\cite{Rosato:2025byu}. As pointed out in~\cite{Rosato:2025byu}, these resonances are characterized by the poles of the greybody factor, corresponding to the long-lived modes. Specifically, the real parts of these QNM spectra determine the positions of the resonance peaks in the greybody factor at low-frequency and in the reflection coefficient at high-frequency, while the imaginary parts are proportional to the resonance width~\cite{Berti:2009wx,Rosato:2025byu}. As for our model, the parabolic-approximated potential also gives rise to long-lived modes and resonance phenomena in the reflection coefficient at high frequencies. However, the greybody factor does not exhibit resonances in the low-frequency regime. That is to say QNM spectra do not provide clear signatures characterizing these resonance behavior, which can be seen from Fig. \ref{R_Log_QNMs}, where the vertical dashed line represents the real part of the QNM spectra. 



The discontinuity in the first derivative of the effective potential also renders some analytic methods, such as the WKB method for computing the greybody factors inapplicable. It is known that the relationship~\cite{Konoplya:2024lir} between the greybody factor $\Gamma_l^{\text{R-W}}(\omega)$ and the fundamental quasinormal mode $\omega_0^{\text{R-W}}$ can be expressed as
\begin{eqnarray}
    \Gamma_l^{\text{R-W}}(\omega)=\Bigg(1+\exp\Big[-\frac{2\pi(\omega^2-\text{Re}(\omega_0^{\text{R-W}})^2)}{4\text{Re}(\omega_0^{\text{R-W}})|\text{Im}(\omega_0^{\text{R-W}})|}\Big]\Bigg)^{-1}+\mathcal{O}(l^{-1})\, .
\end{eqnarray}
The above formula, which is derived from the WKB method, indicates that the position of the greybody factor inflection point is the real part of the fundamental mode. However, this formula does not apply to our model. In fact, we can see from Tab. \ref{QNM_spectra} that the imaginary part of fundamental mode for the piecewise parabolic potential is different from that of the R-W potential, but $\Gamma_l^{\text{Para}}(\omega)$ is close to $\Gamma_l^{\text{R-W}}(\omega)$ as $N$ increases.

Our results clearly demonstrate that the WKB method for computing the greybody factor relies on the smoothness of the effective potential. More importantly, for scenarios where the WKB method becomes inadequate due to non-differentiable effective potentials (particularly in parabolic approximation case), we propose an alternative analytic formulation employing junction conditions through transfer matrix technique. This approach maintains validity for both discontinuous step function potential and piecewise-linear potential configurations. Furthermore, given the robust properties of greybody factors, our proposed method provides a viable framework for approximate calculations of greybody factors, as in the present work.


\section{conclusions and discussion}\label{sec: conclusions}
In this work, we treated the difference between the parabolic approximated potential and the original R-W potential as a perturbation to the latter. We use this framework to investigate the stability of QNM spectra and greybody factors for massless scalar field in the Schwarzschild black hole. In Sec. \ref{sec: parabolic potential}, we construct a piecewise parabolic approximation to the R-W potential and examine its deviation from the latter for different number of segments $N$. Our results (see Fig. \ref{fig_parabolic_approximation} and Fig. \ref{fig_parabolic_approximation_error}) show that the approximated potential systematically converges to the R-W potential as $N$ increases, indicating that the perturbation strength can be effectively controlled by adjusting $N$.

According to the fact that Eq. (\ref{parabolic_potential_equation}) corresponding to the piecewise parabolic potential admits analytic solutions in each  interval, we derive some methods to compute both the QNM spectra and the greybody factors in Sec. \ref{stability_QNMs} and Sec. \ref{sec: GFs} by applying appropriate matching conditions (\ref{QNM_conditions}). Due to the highly nonlinear nature of the condition Eq. (\ref{det_M}), this condition, i.e., $\det(\textbf{M}(\omega))=0$, is not well suited for direct computation of QNMs, but it serves as a useful and reliable tool for validating our numerical results obtained via the so-called multi-domain spectral collocation method. Encouragingly, for the greybody factor, the method proves to be practically effective: using the transfer matrix technique, we obtain fully analytic expressions [see Eq. (\ref{Ain_Aout}) and Eq. (\ref{reflectivity_transmissivity})] for both the greybody factor and the reflection coefficient, even in the presence of non-smooth effective potentials. To our knowledge, this is the first analytic calculation of greybody factors for such non-smooth potentials.

Based on the results presented in Secs. \ref{stability_QNMs} and \ref{sec: GFs}, we find that in our model, QNM spectra exhibit significant instability under small deviations of the R-W potential. Although the parabolic approximated potential becomes increasingly close to the original R-W potential as the number of segments $N$ increases, the corresponding QNM spectra do not show any tendency to converge to the QNM spectra of original R-W case, for either the fundamental mode or the overtones. This is consistent with results of the piecewise step case and the piecewise linear case~\cite{Nollert:1996rf,Daghigh:2020jyk}. In contrast to the QNMs behavior, the greybody factor exhibits remarkably stability. As the difference between the parabolic-approximated potential and the R-W potential decreases, the greybody factors gradually converge to that of the original R-W case (see Fig. \ref{GFs_Plot}). The results for the stability of the greybody factors are consistent with previous related studies~\cite{Oshita:2024fzf,Wu:2024ldo,Rosato:2024arw,Ianniccari:2024ysv}.

Moreover, the QNM spectra of the piecewise parabolic approximated potential exhibit a markedly different asymptotic behavior compared to the standard R-W case. As the overtone number increases, the imaginary parts remain close to zero and become only slightly more negative, indicating slow damping even at high overtones. This is in sharp contrast to the Schwarzschild and other black hole cases, whose imaginary parts rapidly grow in magnitude with overtone number. Such weakly damped overtones are known as long-lived modes and typically appear in ultracompact objects capable of producing gravitational wave echoes~\cite{Shen:2024rup}, such as wormholes. Furthermore, if the greybody factor is analytically continued to the complex frequency plane, the QNM spectra are viewed as poles of the greybody factor. As a result, these long-lived modes are believed to be directly connected to the resonance features observed in the reflection coefficient at low frequency and in the greybody factor at high frequency~\cite{Rosato:2025byu}.

Finally, we emphasize that conventional WKB-based analytic methods for computing the greybody factors are not applicable to our model, due to the non-smooth nature of the effective potential. On the other hand, assuming the general stability of the greybody factor, the analytic method that we proposed based on a transfer matrix technique and piecewise approximations may offer a practical and effective way to compute greybody factors for black holes or other compact objects approximatively.

There are some things that can be further explored among the non-smooth nature. Here, we list some investigations that can be done in the future.
\begin{enumerate}
    \item Despite the frequency-domain instability of QNMs, the prompt time-domain ringdown waveform is stable~\cite{Berti:2022xfj,Spieksma:2024voy,Cao:2024sot,Oshita:2025ibu}. Recently, it is demonstrated that the prompt ringdown waveform remains stable by considering a modified boundary condition and computing the QNM excitation factors~\cite{Glampedakis:2003dn,Oshita:2025ibu}, where the QNM excitation factors are used to reconstruct the waveform~\cite{Oshita:2024wgt}. This inspires us to consider the effect of non-smooth correction of the effective potential on the excitation factor, in order to study the stability of the waveform. Since two amplitudes $A^{\text{in}}(\omega)$ and $A^{\text{out}}(\omega)$ are derived analytically [see Eq. (\ref{Ain_Aout})], excitation factors are able to be calculated directly.
    \item For the Kerr case, one can also study the influences of non-smoothness on QNM spectra and greybody factors, where the carrier of research can be the Sasaki-Nakamura equation~\cite{Sasaki:1981sx} or Teukolsky equation~\cite{Teukolsky:1972my}.
\end{enumerate}


\section*{Acknowledgement}
We are grateful to Yu-Sen Zhou for helpful discussions. This work is supported in part by the National Key Research and Development Program of China Grant No. 2020YFC2201501, in part by the National Natural Science Foundation of China under Grant No. 12475067 and No. 12235019.


\appendix
\section{The analytical solution for a parabolic potential}\label{solution_parabolic_potential}
In this appendix, we solve Eq. (\ref{parabolic_potential_equation}) analytically on an interval $[x_{2k-2},x_{2k}]$. Let $z=tx+s$, i.e., $x=(z-s)/t$, Eq. (\ref{parabolic_potential_equation}) has the following form 
\begin{eqnarray}
    \frac{\mathrm{d}^2\Psi_k}{\mathrm{d}z^2}+\Big[\frac{\omega^2}{t^2}-a\frac{(z-s)^2}{t^4}-b\frac{z-s}{t^3}-\frac{c}{t^2}\Big]\Psi_k=0\, .
\end{eqnarray}
Then, we have
\begin{eqnarray}
    \frac{\mathrm{d}^2\Psi_k}{\mathrm{d}z^2}+\Big[-\frac{a}{t^4}z^2+\Big(\frac{2as}{t^4}-\frac{b}{t^3}\Big)z+\Big(-\frac{as^2}{t^4}+\frac{bs}{t^3}+\frac{\omega^2-c}{t^2}\Big)\Big]\Psi_k=0\, .
\end{eqnarray}
Comparing the Weber equation, we have two transform parameters given by
\begin{eqnarray}\label{functions_s_t}
    t=(4a)^{1/4}\, ,\quad s=\frac{b}{2a}(4a)^{1/4}\, .
\end{eqnarray}
Finally, Eq. (\ref{parabolic_potential_equation}) becomes the standard Weber equation given by~\cite{wang1989special}
\begin{eqnarray}\label{Weber_equation}
    \frac{\mathrm{d}^2\Psi_k}{\mathrm{d}z^2}+\Big(-\frac{z^2}{4}+\nu+\frac{1}{2}\Big)\Psi_k=0\, ,\quad \nu\equiv\frac{b^2}{8a^{3/2}}+\frac{\omega^2-c}{2\sqrt{a}}-\frac{1}{2}\, .
\end{eqnarray}
The two solutions of the above Weber equation are $D_\nu(z)$ and $D_{-\nu-1}(iz)$, which are called the parabolic cylinder functions. Consider that $D_\nu(z)$ and $D_{-\nu-1}(iz)$ are independent, the solution of Eq. (\ref{parabolic_potential_equation}) is given by Eq. (\ref{solution_parabolic_potential_equation}). In addition, as a supplement, we also give the derivative of $\Psi_k$ in terms of $x$. Using the recurrence relation for $D_{\nu}(z)$, i.e., 
\begin{eqnarray}
    \frac{\mathrm{d}D_{\nu}(z)}{\mathrm{d}z}+\frac{z}{2}D_\nu(z)-\nu D_{\nu-1}(z)=0\, ,
\end{eqnarray}
we can compute the derivative $\Psi^{\prime}_k(x)$ of $\Psi_k(x)$ with respect to $x$
\begin{eqnarray}
    \Psi^{\prime}_k(x)&=&C_{1,k} t\Big[-\frac{z}{2}D_{\nu}(z)+\nu D_{\nu-1}(z)\Big]+C_{2,k}(it)\Big[-\frac{iz}{2}D_{-\nu-1}(iz)+(-\nu-1)D_{-\nu-2}(iz)\Big]\nonumber\\
    &=&C_{1,k}t\Big[-\frac{tx+s}{2}D_{\nu}(tx+s)+\nu D_{\nu-1}(tx+s)\Big]\nonumber\\
    &&+C_{2,k}(it)\Big[-\frac{i(tx+s)}{2}D_{-\nu-1}(i(tx+s))+(-\nu-1)D_{-\nu-2}(i(tx+s))\Big]\, ,
\end{eqnarray}
where $k=1,\cdots,N$. Such derivative will be used to solve the QNM spectra and the greybody factor [see Eq. (\ref{QNM_conditions})]. 

\section{The construction of the matrix $\mathbf{M}$}\label{construction_M}
In this appendix, taking $N=3$ as an example, we will explicitly construct the matrix $\textbf{M}(\omega)$, which has been mentioned in Sec. \ref{stability_QNMs}. By applying Eqs. (\ref{QNM_conditions}) on $x_0$, $x_2$, $x_4$ and $x_6$, we have the following equations:
\begin{eqnarray}
    Ae^{-i\omega x_0}&=&C_{1,1}P_1(\omega,x_0)+C_{2,1}P_2(\omega,x_0)\, ,\nonumber\\
    A(-i\omega)e^{-i\omega x_0}&=&C_{1,1}\tilde{P}_1(\omega,x_0)+C_{2,1}\tilde{P}_2(\omega,x_0)\, ,\nonumber\\
    C_{1,1}P_1(\omega,x_2)+C_{2,1}P_2(\omega,x_2)&=&C_{2,1}P_1(\omega,x_2)+C_{2,2}P_2(\omega,x_2)\, ,\nonumber\\
    C_{1,1}\tilde{P}_1(\omega,x_2)+C_{2,1}\tilde{P}_2(\omega,x_2)&=&C_{2,1}\tilde{P}_1(\omega,x_2)+C_{2,2}\tilde{P}_2(\omega,x_2)\, ,\nonumber\\
    C_{2,1}P_1(\omega,x_4)+C_{2,2}P_2(\omega,x_4)&=&C_{2,1}P_1(\omega,x_4)+C_{2,2}P_2(\omega,x_4)\, ,\nonumber\\
    C_{2,1}\tilde{P}_1(\omega,x_4)+C_{2,2}\tilde{P}_2(\omega,x_4)&=&C_{2,1}\tilde{P}_1(\omega,x_4)+C_{2,2}\tilde{P}_2(\omega,x_4)\, ,\nonumber\\
    C_{2,1}P_1(\omega,x_6)+C_{2,2}P_2(\omega,x_6)&=&Be^{i\omega x_6}\, ,\nonumber\\
    C_{2,1}\tilde{P}_1(\omega,x_4)+C_{2,2}\tilde{P}_2(\omega,x_4)&=&B(i\omega)e^{i\omega x_6}\, ,
\end{eqnarray}
where
\begin{eqnarray}\label{P1_P2}
    P_1(\omega,x)&\equiv&D_{\nu}(tx+s)\, ,\nonumber\\
    P_2(\omega,x)&\equiv&D_{-\nu-1}(i(tx+s))\, ,\nonumber\\
    \tilde{P}_1(\omega,x)&\equiv&t\Big[-\frac{tx+s}{2}D_{\nu}(tx+s)+\nu D_{\nu-1}(tx+s)\Big]\, ,\nonumber\\
    \tilde{P}_2(\omega,x)&\equiv&(it)\Big[-\frac{i(tx+s)}{2}D_{-\nu-1}(i(tx+s))+(-\nu-1)D_{-\nu-2}(i(tx+s))\Big]\, ,
\end{eqnarray}
and $s$, $t$, $\nu$ are obtained from Eqs. (\ref{functions_s_t}) and Eq. (\ref{Weber_equation}). It should be emphasized that for the sake of simplicity, we omit the dependency of function $P$ and $\tilde{P}$ on $s$, $t$, $\nu$. In fact, for each interval $[x_{2k-2},x_{2k}]$, three parameters $s$, $t$ and $\nu$ of parabolic cylinder functions are different. Therefore, if the solution vector is written as $[A,C_{1,1},C_{1,2},C_{2,1},C_{2,2},C_{3,1},C_{3,2},B]^T$, the matrix is given by
\begin{eqnarray}
    \mathbf{M(\omega)}=
    \begin{bmatrix}
        -e^{-i\omega x_0} & P_1(\omega,x_0) & P_2(\omega,x_0) & 0 & 0 & 0 & 0 & 0\\
        -(-i\omega)e^{-i\omega x_0} & \tilde{P}_1(\omega,x_0) & \tilde{P}_2(\omega,x_0) & 0 & 0 & 0 & 0 & 0\\
        0 & -P_1(\omega,x_2) & -P_2(\omega,x_2) & P_1(\omega,x_2) & P_2(\omega,x_2) & 0 & 0 & 0\\
        0 & -\tilde{P}_1(\omega,x_2) & -\tilde{P}_2(\omega,x_2) & \tilde{P}_1(\omega,x_2) & \tilde{P}_2(\omega,x_2) & 0 & 0 & 0\\
        0 & 0 & 0 & -P_1(\omega,x_4) & -P_2(\omega,x_4) & P_1(\omega,x_4) & P_2(\omega,x_4) & 0\\
        0 & 0 & 0 & -\tilde{P}_1(\omega,x_4) & -\tilde{P}_2(\omega,x_4) & \tilde{P}_1(\omega,x_4) & \tilde{P}_2(\omega,x_4) & 0\\
        0 & 0 & 0 & 0 & 0 & -P_1(\omega,x_4) & -P_2(\omega,x_6) & e^{i\omega x_6}\\
        0 & 0 & 0 & 0 & 0 & -\tilde{P}_1(\omega,x_4) & -\tilde{P}_2(\omega,x_6) & (i\omega)e^{i\omega x_6}
    \end{bmatrix}\, .
\end{eqnarray}

\section{The analytic expression of the greybody factor for the piecewise parabolic approximation}\label{app: greybody factor}
In this appendix, we are going to solve the analytic expression of the greybody factor for the piecewise parabolic approximated potential. First, we will deal with the case $N=1$. Given $\omega$ and at $x=x_0$, we have two conditions from Eqs. (\ref{QNM_conditions}) given by
\begin{eqnarray}
    e^{-i\omega x_0}&=&C_{1,1}P_1(\omega,x_0)+C_{2,1}P_2(\omega,x_0)\, ,\nonumber\\
    (-i\omega)e^{-i\omega x_0}&=&C_{1,1}\tilde{P}_1(\omega,x_0)+C_{2,1}\tilde{P}_2(\omega,x_0)\, .
\end{eqnarray}
Solving the above linear equation associated with $C_{1,1}$ and $C_{2,1}$, we obtain two functions $C_{1,1}(\omega)$ and $C_{2,1}(\omega)$. The results read
\begin{eqnarray}
    C_{1,1}(\omega)&=&-\frac{e^{-i\omega x_0} \Big[\tilde{P}_2(\omega ,x_0)+i \omega P_2(\omega ,x_0)\Big]}{\tilde{P}_1(\omega ,x_0)P_2(\omega ,x_0)-P_1(\omega ,x_0)\tilde{P}_2(\omega ,x_0)}\, ,\nonumber\\
    C_{2,1}(\omega)&=&-\frac{e^{-i\omega x_0} \Big[\tilde{P}_1(\omega ,x_0)+i \omega P_1(\omega ,x_0)\Big]}{P_1(\omega ,x_0)\tilde{P}_2(\omega ,x_0)-\tilde{P}_1(\omega ,x_0)P_2(\omega ,x_0)}\, .
\end{eqnarray}
Then, at $x=x_2$, two conditions from Eqs. (\ref{QNM_conditions}) are
\begin{eqnarray}
    C_{1,1}(\omega)P_1(\omega,x_2)+C_{2,1}(\omega)P_2(\omega,x_2)&=&A^{\text {in}}(\omega) e^{-i\omega x_2}+A^{\text {out}}(\omega) e^{+i\omega x_2}\, ,\nonumber\\
    C_{1,1}(\omega)\tilde{P}_1(\omega,x_2)+C_{2,1}(\omega)\tilde{P}_2(\omega,x_2)&=&A^{\text {in}}(\omega)(-i\omega)e^{-i\omega x_2}+A^{\text {out}}(\omega)(i\omega)e^{+i\omega x_2}\, .
\end{eqnarray}
Solving the above equation, one gets
\begin{eqnarray}
    A^{\text{in}}(\omega)&=&\frac{e^{i\omega x_2 }}{2 \omega }\Big[C_{1,1}(\omega)\omega  P_1(\omega,x_2)+i C_{1,1}(\omega) \tilde{P}_1(\omega ,x_2)+C_{2,1}(\omega)\omega  P_2(\omega ,x_2)+i C_{2,1}(\omega)\tilde{P}_2(\omega ,x_2)\Big]\, ,\nonumber\\
    A^{\text{out}}(\omega)&=&\frac{e^{-i\omega x_2}}{2\omega}\Big[C_{1,1}(\omega)\omega P_1(\omega ,x_2)-i C_{1,1}(\omega)\tilde{P}_1(\omega ,x_2)+C_{2,1}(\omega)\omega  P_2(\omega ,x_2)-iC_{2,1}(\omega)\tilde{P}_2(\omega ,x_2)\Big]\, .
\end{eqnarray}
Finally, the greybody factor can be derived from Eqs. (\ref{reflectivity_transmissivity}). In fact, we can use the matrix representation to give $A^{\text{in}}(\omega)$ and $A^{\text{out}}(\omega)$ for the general case. Following the method of $N=1$ mentioned above, it is not difficult to obtain the general result. In terms of methodology, it means continuously applying junction conditions (\ref{QNM_conditions}). Therefore, given $\omega$, for general $N$, solving $A^{\text{in}}(\omega)$ and $A^{\text{out}}(\omega)$ analytically, we have
\begin{eqnarray}\label{Ain_Aout}
    \begin{bmatrix}
        A^{\text{in}}(\omega)\\
        A^{\text{out}}(\omega)
    \end{bmatrix}&=&
    \begin{bmatrix}
        e^{-i\omega x_{2N}} & e^{i\omega x_{2N}}\\
        (-i\omega)e^{-i\omega x_{2N}} & (i\omega)e^{i\omega x_{2N}}
    \end{bmatrix}^{-1}\nonumber\\
    &&\bm{\cdot}
    \Bigg\{\begin{bmatrix}
        P_1(\omega,x_{2N}) & P_2(\omega,x_{2N})\\
        \tilde{P}_1(\omega,x_{2N}) & \tilde{P}_2(\omega,x_{2N})
    \end{bmatrix}\bm{\cdot}
    \begin{bmatrix}
        P_1(\omega,x_{2N-2}) & P_2(\omega,x_{2N-2})\\
        \tilde{P}_1(\omega,x_{2N-2}) & \tilde{P}_2(\omega,x_{2N-2})
    \end{bmatrix}^{-1}\Bigg\}\bm{\cdot}\cdots\nonumber\\
    &&
    \bm{\cdot}\Bigg\{\begin{bmatrix}
        P_1(\omega,x_{2}) & P_2(\omega,x_{2})\\
        \tilde{P}_1(\omega,x_{2}) & \tilde{P}_2(\omega,x_{2})
    \end{bmatrix}\bm{\cdot}
    \begin{bmatrix}
        P_1(\omega,x_{0}) & P_2(\omega,x_{0})\\
        \tilde{P}_1(\omega,x_{0}) & \tilde{P}_2(\omega,x_{0})
    \end{bmatrix}^{-1}\Bigg\}\bm{\cdot}
    \begin{bmatrix}
        e^{-i\omega x_0}\\
        (-i\omega)e^{-i\omega x_0}
    \end{bmatrix}\, ,
\end{eqnarray}
where $\cdot$ denotes matrix multiplication, and $[\cdot]^{-1}$ denotes inverse of matrix. Here, four functions $P_1(\omega,x)$, $P_2(\omega,x)$, $\tilde{P}_1(\omega,x)$ and $\tilde{P}_2(\omega,x)$ are given by Eqs. (\ref{P1_P2}). Note that the number of the term $\{\cdot\}$ is $N$. According to the definition of the QNM spectrum, one substitutes the data in the Tab. \ref{QNM_spectra} into equation (\ref{Ain_Aout}) and finds that $A^{\text{in}}(\omega)=0$ up to the numerical error.

\bibliography{reference}
\bibliographystyle{apsrev4-1}

\end{document}